\documentclass[
    ,final            
  ]
  {aipproc}

\layoutstyle{8x11single}

\begin{document}

\title{Recent Progress in Applying Gauge/Gravity Duality to Quark-Gluon Plasma Physics}


\author{Andreas Karch}{
  address={Department of Physics, University of Washington, Seattle, Wa 98195, USA}
}

\begin{abstract}
We give a brief overview of the basic philosophy behind applying gauge/gravity duality  to problems in heavy ion physics. Recent progress regarding viscosities, anomalous hydrodynamics, energy loss, as well as thermalization will be reviewed.
\end{abstract}

\maketitle


\section{Introduction}
Gauge/gravity duality (or ``holography" for short) provides us with many examples of solvable toy models of strong coupling dynamics. Solvable models are useful whenever the real problem is too hard to tackle. Latter is unfortunately always the case when strong coupling, such as it is present in quantum-chromodynamics (QCD) at the energies explored in heavy ion collisons at RHIC or at the LHC, meets either real time dynamics or finite density. For such processes the standard tool to analyze strongly coupled processes, the lattice, has difficulties. Lattice techniques allow to evaluate the formal path integral that defines any quantum field theory by approximating it by a finite number of ordinary integrals. A path integral can be viewed as an integral over the values the fields can take at any point in space and time. By replacing space-time with a finite lattice the number of points in space-time is actually finite and so the path integral simply becomes a finite set of ordinary integrals. In practice these are still too many integrals even for a computer to do in a realistic amount of time, so one has to rely on importance sampling. If one is interested in the Euclidean path integral, as appropriate for equilibrium physics, the $e^{- S}$ weight in the path integral tells us that one can approximate the integrals by a sum over a small subset of configurations as long as one ensures that the most important configurations, that is those with the least action $S$, are captured. For real time processes with their $e^{i S}$ weight or processes with finite baryon density, where the Euclidean action is typically complex, one basically needs to sample the full configuration space\footnote{There has been some recent progress on calculating real-time transport properties, such as the viscosity, using lattice methods, see e.g. \cite{Meyer:2011gj} for a review. If one were to know the Euclidean physics and in particular the spectral functions (that is the Euclidean 2-pt correlation functions) analytically, one could recover hydrodynamic transport properties from analytic continuation. As long as the Euclidean correlators are only known in a numerical approximation, one needs to rely on templates for the functional form of the spectral function to which the numeric data can be fitted. This comes with substantial systematic uncertainties -- the form of the template strongly influences the final answer one obtains.}.

Examples of physical situations where real time physics is important is the study of the evolving fireball in heavy ion-collisions\footnote{Another area of physics that requires at least a qualitative understanding of strong coupling physics is the study of condensed matter systems that defy Landau's paradigms -- non-Fermi liquids and phase transitions that are not simply governed by an effective potential for the order parameter. There has been a lot of recent activity applying holography to these questions, but this is for a different audience.}, which involves both far-from-equilibrium physics (the early stage thermalization) as well as near-equilibrium-physics (the hydrodynamic evolution of the fireball). Before the advent of holographic models, the most commonly used theoretical tool to analyze these processes was perturbation theory. Perturbation theory certainly applies to QCD at asymptotically high temperatures. To describe QCD at RHIC temperatures one can extrapolate to strong coupling by using the leading order perturbative result and simply plugging in values of the coupling of order 1 into the resulting formulas. This procedure would lead one to believe that around the transition temperature the quark gluon plasma should have a fairly large viscosity to entropy ratio and hard probes in this plasma should only experience relatively modest energy loss. Both these expectations have been blown out of the water by the experimental realities at RHIC. Quarks and gluons seem to form a strongly correlated liquid rather then a plasma; transport properties often differ by an order of magnitude from naive perturbative extrapolations. This is the perfect arena for holographic toy models.

In holographic toy models one can get exact answers for transport properties, alas only in the theories that allow a solvable holographic description which unfortunately do not include QCD. So instead of calculating transport in the correct theory but in the wrong regime as one did in the perturbative extrapolation, one is calculating in the right regime but in the wrong theory. As such, one should expect errors of the order of 100\% and these are all systematic -- there is no small expansion parameter that systematically corrects for calculating with the wrong theory. An honest theorist should be ashamed of using either approach. Still, a factor of 2 mismatch with reality could be seen as a success when compared to the order of magnitude mismatch of extrapolated perturbation theory. But much more than that, holographic modeling seems to give us a better intuition about the microscopic mechanism that underly transport in a strongly coupled theory. At weak coupling, the microscopic description of the system is always in terms of quasi-particles that travel a mean free path larger than their typical size between collisions. While details clearly depend on the nature of the quasi-particles and their interactions, qualitative features are basically the same in all weakly coupled systems. Holography is the best tool we have right now to qualitatively tell us how a system behaves when the quasi-particle picture breaks down, as it seems to do in the quark-gluon liquid produced at RHIC and LHC. Instead of an assembly of quasi-particles, the best way to think of a strongly coupled liquid appears to be its representation as a five dimensional black hole.

Before describing some of the recent qualitative insights that have been gained using this holographic paradigm (and have clearly had an impact on how many theorists and experimentalists think about the quark-gluon liquid), let us at least say a few words about what the models are that can be solved using these methods. By now a very large class of holographically solvable theories has been found. This includes theories in 1, 2, 3, 4, 5, or 6 space-time dimensions
(any dimension one could possibly care about experimentally and then some), with our without super-symmetry,
conformal theories or theories that confine and produce a mass gap,
theories with or without chiral symmetry breaking. All these theories can be studied
at finite temperature and density. The unifying features that are however common to all these theories is that they are\footnote{In some special examples of holographic theories, e.g. the M2 and M5 brane theories, there is a single parameter whose largeness ensures both of the properties we describe here: classicality and separation of scales. They are still both essential for a simple holographic description.} at ``large $N$" and ``large $\lambda$". "Large $N$" means that the theory has a large number of degrees of freedom; in examples that involve non-Abelian gauge theories $N$ typically is the number of colors. The large $N$ limit ensures that the theory essentially behaves classically. While $N=3$ in QCD it is known that for many questions large $N$ may actually not be a bad approximation. ``Large $\lambda$" implies that the theory is not just at couplings of order 1 but actually at parametrically strong coupling (in a non-Abelian gauge theory $\lambda$ is the standard `t Hooft coupling). What this means in practice is that there is a large separation of scales in the spectrum. Take for an example a confining holographic theory\footnote{The corresponding argument for a conformal theory involves the spectrum of operator dimensions instead of the particle spectrum.}. Due to large $\lambda$ in this theory there should be a large separation between the mass of the lightest spin 1 meson (whose mass is set by the confinement scale, $\Lambda_{QCD}$) and the lightest spin 2 meson (whose mass would be set by the ``string tension", and so would scale as $\lambda^{1/4} \Lambda_{QCD}$). Confronting this with QCD one sees that the corresponding mesons weigh in at 775 MeV and 1275 MeV respectively: clearly both are governed by the same mass scale. QCD does not have a simple holographic dual.

\section{Hydrodynamics}

Hydrodynamics is the universal theory describing long-wavelength fluctuations of any near-equilibrium system, including the fireball at RHIC. Like any low-energy effective theory, it is governed by a small set of transport coefficients that need to be extracted from the microscopic theory or model. The most prominent of these transport coefficients is the (shear) viscosity\footnote{As we will not consider bulk viscosity in this review, we will from now on simply refer to the shear viscosity as ``the viscosity".}. It's relative smallness in the quark-gluon liquid is what ensured the tremendous success of hydrodynamic modeling of the expanding plasma.

\subsection{Viscosities}

While of course no longer a ``recent" discovery, no review of applications of holographic methods to heavy-ion physics would be complete without a discussion of the viscosity. To appreciate why this topic has received so much attention, let us briefly recapitulate a few facts about the viscosity. In a fluid experiencing a shear (a velocity gradient in the direction transverse to the flow direction) a given fluid layer experiences a force per unit area proportional to the velocity gradient. The constant of proportionality is called the viscosity; as such it has units of force/(area $\cdot$ velocity/length),
that is in SI-units it is measured in $Pa \cdot s$. A more commonly unit used to express viscosities is the centipoise ($cP$) where $1 cP = 10^{-3} Pa \cdot s$. The reason that the centipoise is popular is that the most well known fluid to mankind, water, has a viscosity of about 1 $cP$. In comparison, the viscosity of air is about 0.02 $cP$ whereas the viscosity of honey is anywhere between 2000 $cP$ and 10000 $cP$ (depending on brand). For most fluids, measuring viscosity is fairly straightforward. But sometimes it can also be somewhat challenging: let us take a look at two experiments that had to overcome significant obstacles to get a quantitative handle on the viscosity of their fluid of interest. The first experiment is the famous pitch drop experiment performed at the University of Queensland in Brisbane, Australia. Many facts about this experiment can easily be found on Wikipedia. The goal of the experiment is to measure the viscosity of pitch by measuring the speed at which it ``flows" out of a funnel. It's fame is not necessarily based on the fundamental interest of the physics community in pitch, but rather on the fact that it is officially the world's longest running laboratory experiment. The experiment was started in 1927; during it's 8 decades of running a total of 8 drops of pitch have fallen out of the funnel\footnote{Apparently no one has ever witnessed a drop fall; the webcam installed after drop number 7 unfortunately malfunctioned for drop number 8.}. From the data collected so far one can estimate the viscosity of pitch to be about $2.3 \times 10^{11}$ $cP$, 230 billions times that of water. For their efforts John Mainstone, the current custodian of the experiment, and the late Thomas Parnell, who originally started the experiment in 1927, were awarded the 2005 Ig Nobel Prize in Physics.

The value of $2.3 \times 10^{11}$ $cP$ for the viscosity of pitch is something to keep in mind when thinking about another famous experiment that measured viscosity: from the hydrodynamic modeling of the fireball produced at RHIC one can extract the viscosity of the quark-gluon liquid. It's viscosity is about $10^{14}$ $cP$, almost three orders of magnitude larger than the viscosity of pitch. Nevertheless, while the former value seems to be somewhat of a curiosity and is typically received in a somewhat humorous fashion, the latter was announced by BNL with a press release stating that
``the degree of collective interaction, rapid thermalization and {\it extremely low viscosity} [[italics added for emphasis]] of the matter being formed at RHIC makes this the most nearly perfect liquid ever observed". One of the main motivations behind this at first sight somewhat peculiar interpretation of the value measured at RHIC is a calculation performed using holographic toy models. Following an earlier calculation in Ref. \cite{Policastro:2001yc} in one particular holographic model, it was argued by Kovtun, Son and Starinets (KSS) \cite{Kovtun:2004de} that in a large class of holographic models the shear viscosity $\eta$  is given by:
$$ \frac{\eta}{s} = \frac{\hbar}{4 \pi} $$
were $s$ stands for the entropy density of the fluid. The same quantity grows as an inverse power of the coupling at weak coupling. It is this quantity, $\eta/s$, that in fact is smaller for the QCD plasma than for any other fluid known to mankind (cold atomic gases are the only other fluid that comes close). In fact, most current experimental data puts $\eta/s$ for the quark-gluon liquid in the vicinity of the KSS value. For comparison, $\eta/(\hbar s)$ is about 30 for water and still about 10 for superfluid helium, whereas\footnote{For pitch no reliable value for its entropy density seems to be available in the standard tables; but presumably it is not that different from that of water, so there is little doubt that $\eta/s$ for pitch is orders of magnitude larger than for water.} $1/(4 \pi) \approx 0.08$.

The holographic calculation here was crucial in many ways. For one, it pinned down the correct observable to compare between fluids. Of course practitioners knew long ago that the important dynamical quantity is the force to mass ratio (which after all sets the acceleration) and not the force itself. But establishing $\eta/s$ as the crucial quantity to measure is certainly one thing holography has contributed to our understanding of heavy ion physics. Holography also gave us a benchmark. It is by now standard operating procedure that all experimental talks on viscosities compare the results they get to the universal holographic value. However there is one thing $\eta/s = \hbar/(4 \pi)$ is not: a bound.

The original KSS paper observed that $\eta/s$ goes to infinity a weak coupling and in a large class of infinite coupling theories goes to $\hbar/(4 \pi)$. This pattern suggests that potentially $\hbar/(4 \pi)$ is an absolute lower bound: as it is reached at infinite coupling one may suspect that $\eta/s$ always is above this value for any large but finite coupling. We know by now that this is not the case. The most precise counterexample is provided by a holographic calculation by Kats and Petrov \cite{Kats:2007mq} who found that an ${\cal N}=2$ supersymmetric gauge theory with a $Sp(N)$ gauge group, a matter hypermultiplet in the 2-index anti-symmetric tensor representation as well as 4 hypermultiplets in the fundamental representation has
$$ \frac{\eta}{s} = \frac{\hbar}{4 \pi} \left ( 1 - \frac{1}{2 N} + {\cal O} (1/N^2) \right ).$$
As terms of order $1/N^2$ are neglected, this calculation is only valid in the large $N$ limit. So while the putative bound is violated, it is not violated by much. The difference in this example compared to the set of theories considered by KSS is that the holographic dual is not simply a gravitational theory dominated by the Einstein-Hilbert term, but in addition includes curvature squared terms. In the large $N$ limit these curvature squared terms are the leading correction to the Einstein-Hilbert answer and they violate the bound. Terms containing even higher powers of the curvature terms are suppressed by additional powers of $N$. It has become clear that higher curvature terms quite generically can violate the bound \cite{Kats:2007mq,Brigante:2007nu,Buchel:2008vz}. Truncating the theory to include a finite number of higher derivative terms, $\eta/s$ can in principle be made as small as one wants. However typically some problems (for example violations of causality) arise beyond a certain value of $\eta/s$ between 0 and $\hbar/(4\pi)$. The precise number depends however on the details. Of course neglecting corrections which involve even larger powers of the curvature is only justified whenever the corrections to $\eta/s$ implied by the finitely many higher curvature terms one kept are small. So in a controlled holographic setting the violations of the bound are always small. But this seems to be a rather technical limitation. These results aren't really recent anymore -- the Petrov and Kats paper is already 4 years old. The biggest news here is there is no news. We all like our viscosity bounded and the fact that the counterexamples still stand despite ongoing efforts to make them go away makes it likely that these results are here to stay.

\subsection{Second Order Hydro}
Viscosity is not the only hydrodynamical transport coefficient that
can be calculated holographically. Viewing hydrodynamics as a long-distance approximation it is clear that the viscosity, which gives a force as response to a velocity gradient is only the first in a series of terms proportional to higher spatial derivatives of the velocity field. Of course for this expansion to be well defined, one needs that the terms containing two derivatives ("2nd order hydro") are negligible compared to the viscosity term. While true in principle, for practical purposes 2nd order hydro terms need to be included in numerical simulations. Viscous hydrodynamics without the 2nd order terms becomes acausal at short distances. Conceptually, this is not a problem. Hydrodynamics is no longer a valid approximation of nature at very short distances. However, to numerically simulate the evolution of the fireball one needs to cut off this acausal behavior. Fortunately, simply including the 2nd order terms does this job. In the end one then needs to check that while the inclusion of 2nd order hydro terms is crucial for numerical stability, the actual values do not significantly impact physical predictions. This is typically the case for simulations of the RHIC firelball. 2nd order hydro coefficients for holographic toy models were calculated in 2007 in Refs. \cite{Baier:2007ix} and \cite{Bhattacharyya:2008jc}. The results gave a systematic classification of all allowed 2nd order terms. They also gave benchmark values, which are routinely used in hydro simulations. One takes the holographic values as a starting point and makes sure that varying them by factors of 2 or so has no impact on the outcome. But of course changing them by orders of magnitude will at some point alter the physical results, so having a rough estimate of a ballpark figure as provided by holography is exactly what is needed.

\subsection{Anomalous Hydro}

The most recent major development in applying holography to hydrodynamics of strongly coupled liquids is the systematic understanding of novel hydrodynamic transport coefficients driven by anomalies in the underlying quantum field theory. Let us start with stating the results, which are most succinctly summarized in a recent work by Kharzeev and Son \cite{Kharzeev:2010gr}. For any conserved charge, there exists an anomaly dominated contribution to the corresponding current $\vec{J}$ which is given by
$$ \vec{J} = \frac{N \mu_5}{2 \pi^2} \left [ tr(VAQ) \vec{B} + tr(VAB) 2 \mu \vec{\omega} \right ]$$
$N$ is the number of colors, $N=3$ in the real world. The two conserved charges of interest in heavy-ion collisions are electric charge or baryon number. The above formula applies to either $\vec{J}$ being the electric or the baryon number current. The first term gives us a current
in response to a magnetic field $\vec{B}$; this is known as the ``Chiral Magnetic Effect"(CME). The second term gives a similar current in response to the vorticity of the the flow (that is the curl of velocity field); this is known as the
``Chiral Vortical Effect" (CVE). Both terms are proportional to $\mu_5$, the axial chemical potential. Non-zero $\mu_5$ implies an imbalance between left-handed and right-handed fermion numbers. In heavy ion collisions one starts with a symmetric initial state, so there will be, on average, no such imbalance in the final state: $\langle \mu_5 \rangle=0$. Due to event-by-event fluctuations in a given event there can be regions with some net axial charge, $\langle \mu_5 \rangle \neq 0$. The magnitude of these fluctuations is uncertain and it is a topic of current experimental interest to determine whether observed charge asymmetries are due to this effect. What makes these effects however so remarkable from the theoretical point of view is that the coefficients, $tr(VAQ)$ and
$tr(VAB)$ are determined by anomalies. Their one-loop contribution arising from triangle diagrams is the exact answer. For three flavors the coefficients in the CME are 2/3 and 0 for electric and baryon number current respectively, they are 0 and 1 for the CVE. As the CVE is also proportional to the chemical potential of the conserved charge itself and since this decreases with increasing $\sqrt{s}$, one can arrive at a set of generic predictions \cite{Kharzeev:2010gr}: charge asymmetries have to be accompanied by same sign baryon number asymmetries and the ratio between baryon and charge asymmetry should increase
as $\sqrt{s}$ is lowered.

While the connection between CME and anomalies had been appreciated before, the connection between the CVE and anomalies is a little bit more surprising. The first quantitative connection between the CVE and the triangle anomaly has appeared in a holographic calculation. In Refs. \cite{Erdmenger:2008rm} and \cite{Banerjee:2008th} a CVE contribution to a conserved current in a holographic fluid has been identified. In these particular calculations this contribution was entirely due to ``Chern-Simons-terms", the holographic manifestation of the field theory anomalies. While in principle it could have been the case that this connection between anomalies and CVE was an infinite coupling artefact, it was suspected that the connection is probably deeper and true at all couplings. In fact, in a remarkable paper Son and Surowka \cite{Son:2009tf} have shown that the CVE can be derived from the anomaly with the additional input that the entropy current in hydrodynamics always has to have a non-negative divergence (so that entropy never decreases). In the presence of external fields (which drive currents due to the anomaly) positivity of the divergence of the entropy current can only be maintained if the CVE with a fixed coefficient is taken into account. While this proof itself does not use holography, it was certainly important to know the right answer before setting out to give a proof that is valid beyond the narrow regime of validity of the holographic approach.

\section{Energy Loss}

In addition to the evidence form hydrodynamics, the other large class of observables that tell us that the quark-gluon liquid produced at RHIC and LHC is strongly coupled involve the study of hard particles (that is particles with energy much larger than the temperature) traveling through the plasma. For some subset of events, such hard probes are produced in the same collision that set off the fire ball to begin with. They are the result of head-on collisions of two partons in the colliding nuclei.

In the holographic setting, the easiest hard probe to analyze is a heavy quark \cite{Herzog:2006gh,CasalderreySolana:2006rq,Gubser:2006bz}. A heavy quark loses momentum $p$ at a rate dominated by a friction coefficient $\mu$:
$$ \frac{dp}{dt} = - \mu p, \quad \quad \mu= \frac{\frac{\pi}{2} \sqrt{\lambda} T^2}{m}$$
The fact that the loss rate is inversely proportional to the mass of the quark is quite generic in holographic models and distinct from weak coupling extrapolations. If both charm and bottom quarks behave in this fashion the ratio of their suppression factors should reflect this one over mass dependence, see e.g. \cite{Ficnar:2011yj}. Theoretically maybe more interesting is the fact that the {\it mechanism} for energy loss is qualitatively different from weak coupling. At weak coupling hard probes lose energy by radiating off order one fractions of their energy and momentum in a single emission event. In contrast, the holographic heavy quark drag seems to be best interpreted as due to the emission of very many very soft gluons. To the extend that the latter is the right dynamical picture for energy loss in QCD around the temperatures probed in heavy ion collisions we expect this to be one more case where the weakly coupled quasiparticle picture has to be replaced by the intuition gained from holography.

\begin{figure}
  \includegraphics[height=.24\textheight]{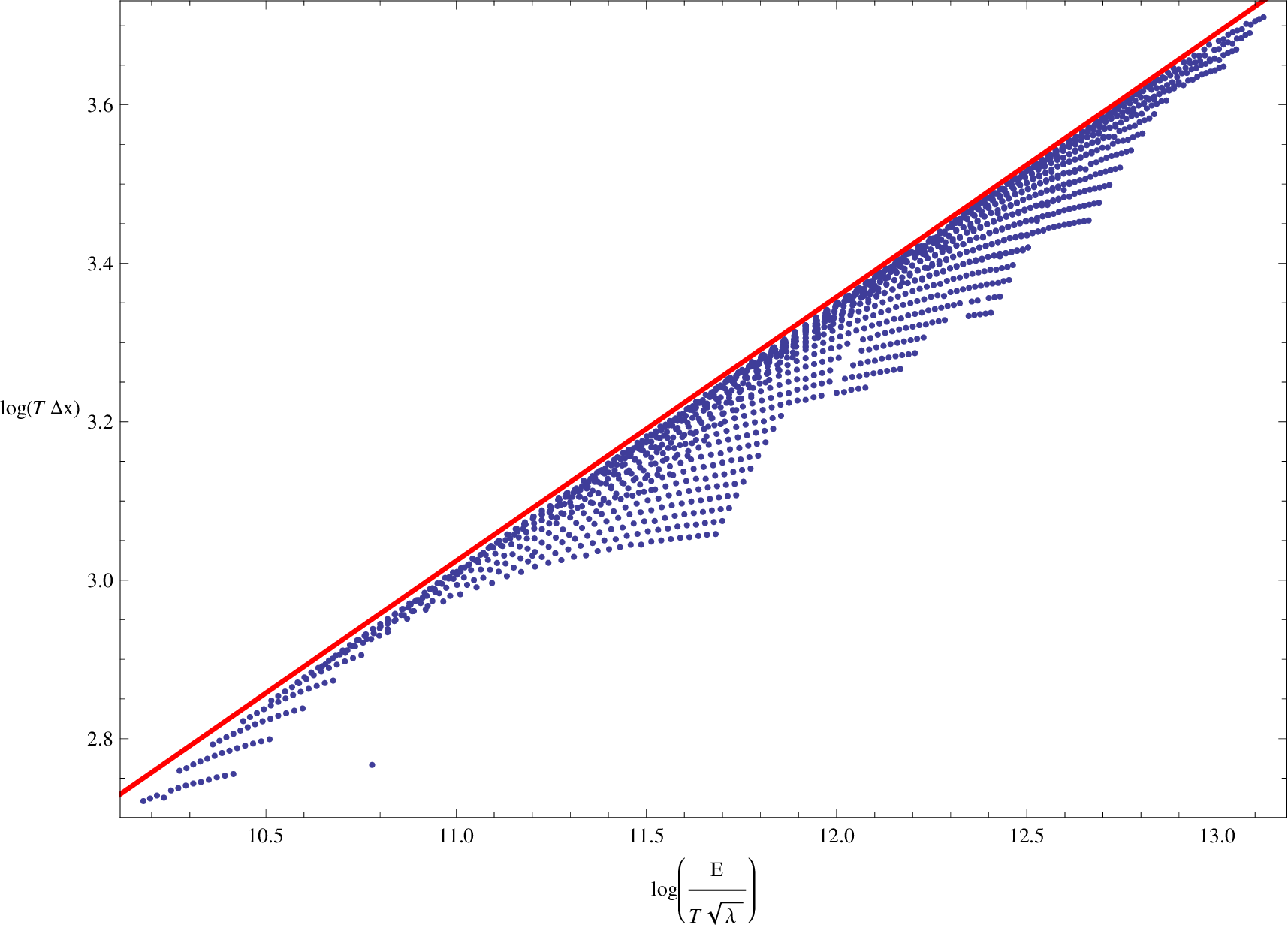}
  \caption{Stopping Distance versus energy for a collection of different initial conditions corresponding
  to a quasi-particle traveling through the holographic plasma.}
  \label{eplot}
\end{figure}

More recently energy loss has also been determined for light quarks and gluons \cite{Chesler:2008uy,Gubser:2008as}. These excitations are more subtle as they are not fundamentally different from the majority of particles the plasma is made from. As long as they have an energy much larger than the typical thermal fluctuation in the plasma, they act as a quasiparticle which loses energy to the surrounding plasma. But once their energy becomes of order the temperature, they just become one with the plasma, their energy and momentum is thermalized and diffuses away. Unlike the heavy quark drag, the corresponding loss rates can not be extracted from a stationary process; the holographic dual of the light quark falls into the higher dimensional black hole that is the holographic dual of the plasma. A common way to characterize the energy loss by a light quark is in terms of its {\it stopping distance}, that is the total distance the probe travels from its creation to its eventual thermalization in the plasma. For a weakly coupled probe, this stopping distance grows with energy\cite{Baier:1996vi,Baier:1996kr} as $E^{1/2}$. This includes the case \cite{Liu:2006ug} where the plasma itself is strongly coupled, but the probe's energy loss rate is still dominated by a few bremsstrahlung emissions enabled by the plasma.

To address the question of stopping distance in the holographic model, we numerically simulated thousands of string configurations falling into the black hole in \cite{Chesler:2008uy}. We also were able to obtain a good analytic approximation in the limit that the string initially carries much more energy than the typical thermal fluctuation. The stopping distance and energy for every single one of these simulations are plotted in Fig. \ref{eplot}. The most important lesson that can be extracted is that the {\it maximal} stopping distance a light quark can travel for a given energy scales as $E^{1/3}$, not $E^{1/2}$ (the red line in the plot above). However the details depend on the exact initial condition the strings is started out with. Recall that in the strongly coupled plasma quark dynamics is governed by the rapid emission of many soft quanta. Correspondingly the hard probe should not simply be thought of as a quark, but rather as a quasi-particle consisting of a quark and a cloud of soft gluons around it. The state of the cloud has a significant impact on the distance the quark can travel.

In more recent work \cite{Arnold:2010ir} Arnold and Vaman considered jets created by a single off-shell photon hitting the holographic plasma. In that case they can quantify how likely a quark/anti-quark pair with a particular gluon cloud is produced. By calculating a holographic 3-pt function corresponding to observing the charge deposition in the jet state created by the off-shell photon they were able to confirm our scaling of the maximal stopping distance with $E^{1/3}$, but in addition identified that the most likely distance traveled by a quark of energy $E$ scales as $E^{1/4}$.

While it is of course not clear in detail which, if any, of these properties are shared by hard probes in the QCD quark-gluon liquid, what is reassuring is that the qualitatively different loss mechanism at strong coupling does not just give rise to different numerical coefficients, but to completely different {\it exponents}. One could hope that these scaling laws correspond to universal properties of strongly coupled liquids\footnote{Of course one should be very cautious when claiming to deduce universal principles from one class of examples.}. With the wealth of new data coming in both from RHIC and the LHC the experimental understanding of energy loss should significantly improve in the month or at least years to come and hopefully we will be able to tell what power law is a better fit to the corresponding processes in the real world.

\section{Thermalization}

Another big puzzle in heavy ion collisions is the seemingly rapid thermalization of the fireball. Once more, estimates based on perturbation theory extrapolated to couplings of order one seem to be in sharp contrast to the experimental reality. So how quickly does the holographic fireball thermalize? To answer this question, one wants to study in the holographic model a system initially at zero temperature perturbed by an external force at $t=0$. The perturbation dumps energy into the system and one wants to watch the response. Thermalization in the field theory corresponds to the formation of a black hole in the holographic dual. Different versions of this story have been studied over the past 3 years, the most recent and most closely related to heavy-ion physics is a numerical study by Chesler and Yaffe \cite{Chesler:2010bi}. In this work the authors set up two thin slices of energy density, localized in the $z$ direction but infinitely extended in the transverse $x$-$y$ space. They then send these two shockwaves at each other at the speed of light and observe what happens when they collide. The infinitely extended shocks are thought to be a reasonable approximation to the energy density carried by the pancake shaped gold or lead nuclei of real heavy-ion collisions.

\begin{figure}
  \includegraphics[height=.2\textheight]{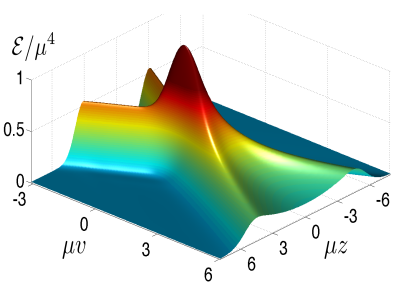}
  \caption{Profile of the energy density corresponding to two colliding shockwaves. $z$ is the direction along the beampipe, whereas $v$ is time. $\mu$ is an energy scale that sets the initial energy density in the shocks.}
  \label{cy}
\end{figure}

The energy density as a function of position $z$ and time $v$ is displayed in figure Fig \ref{cy}, which is taken from Ref. \cite{Chesler:2010bi}. For $v<0$ we see the two shocks moving along the ``beampipe" (the $z$-direction) at the spead of light (at 45 degrees). They collide around $v=0$. After the sharp spike in energy density at the collision moment one sees two diminished shocks (the remnants of the ``nuclei") continue along the beampipe. The energy density between those outgoing shocks is the expanding quark-gluon plasma. The shocks used are characterized by two scales, $\mu$, which sets the initial energy density in the shockwaves and $w$, which sets the width (the combination of $w$ and $\mu$ then uniquely determines the height). On dimensional grounds, $z$, $v$ and $w$ are measured in units of $\mu$. In their published work, Chesler and Yaffe show results for $w=0.75/\mu$, close to what is thought to be appropriate for RHIC\footnote{According to private conversations with the authors, as the width increases the fireball takes longer to thermalize, as the ongoing collision process leads to continuous production of non-equilibrated energy density.}.
 
The holographic dictionary allows one to extract the full expectation value of the stress tensor for this collision, in particular the longitudinal and transverse pressure. Energy densities and pressure can be compared to hydrodynamic evolution. Chesler and Yaffe find that, while around $v=0$ the collision is not well approximated by hydro, the agreement sets in at times around $2/\mu$. At this stage the system seems to have reached local equilibrium (or at least isotropy).
If one defines the thermalization time from the time the shocks start to overlap (which is before $v=0$), this corresponds to a thermalization time of about $4/\mu$. To give a crude comparison to what this may imply for RHIC, Chesler and Yaffe estimate the energy density in the gold nuclei to correspond to a $\mu$ of about 2.3 GeV. This gives a thermalization time of about 0.35 fm/c. This is consistent with experiment as it is well below the 1 fm/c that is often taken as the latest thermalization time compatible with RHIC data. Clearly rapid thermalization is possible in strongly coupled systems. The really important question now seems to be what else to ask from these simulations. Chesler and Yaffe solved a hard numerical general relativity problem; their solution contains a lot of extra information about the thermalization process. What are the right questions to ask? Can one learn generic lessons?

\section{Discussion}
Holographic models provide us with qualitative insights into strong coupling dynamics. I reviewed three areas in which these studies have clearly benefitted our theoretical understanding, all three relevant to heavy-ion physics: hydrodynamics, energy loss, and thermalization. One large body of work that hasn't been touched upon in this review are the ongoing studies of meson and baryon masses and matrix elements using holographic techniques. For these static questions, we have a great tool (the lattice) and it is unlikely that holography will ever be competitive. However there is still a great opportunity for holography when it comes to studying similar properties of strongly coupled theories {\it other} than QCD. Many models of beyond the standard model physics require a new, strongly coupled sector. Unlike for nuclear forces in this case we do not know yet which one is the right Lagrangian. One needs to be able to quickly work out properties of many strongly coupled theories to see which one can fit the (hopefully soon) existing data. A quick-and-dirty model as provided by holography is well suited for this explorative stage.


\begin{theacknowledgments}
I'd like to thank the organizers of the PANIC11 conference for inviting me and Paul Chesler, Chris Herzog, Kristan Jensen, Can Kozcaz, and Larry Yaffe for the pleasant collaborations on some of the work reviewed in here. In addition I'd like to thank my colleague Dam Son for very useful discussions on his work and for inspiration for this review. This work was supported in part by the U.S. Department of Energy under Grant No. DE-FG02-96ER40956.
\end{theacknowledgments}



\bibliographystyle{aipproc}   

\bibliography{Panic}

\begin{thebibliography}{23}
\expandafter\ifx\csname natexlab\endcsname\relax\def\natexlab#1{#1}\fi
\providecommand{\enquote}[1]{``#1''}
\expandafter\ifx\csname url\endcsname\relax
  \def\url#1{\texttt{#1}}\fi
\expandafter\ifx\csname urlprefix\endcsname\relax\def\urlprefix{URL }\fi
\providecommand{\eprint}[2][]{\url{#2}}

\bibitem[Meyer(2011)]{Meyer:2011gj}
H.~B. Meyer, \emph{Eur. Phys. J.} \textbf{A47}, 86 (2011), \eprint{1104.3708}.

\bibitem[Policastro et~al.(2001)]{Policastro:2001yc}
G.~Policastro, D.~Son, and A.~Starinets, \emph{Phys.Rev.Lett.} \textbf{87},
  081601 (2001), \eprint{hep-th/0104066}.

\bibitem[Kovtun et~al.(2005)]{Kovtun:2004de}
P.~Kovtun, D.~Son, and A.~Starinets, \emph{Phys.Rev.Lett.} \textbf{94}, 111601
  (2005), an Essay submitted to 2004 Gravity Research Foundation competition,
  \eprint{hep-th/0405231}.

\bibitem[Kats and Petrov(2009)]{Kats:2007mq}
Y.~Kats, and P.~Petrov, \emph{JHEP} \textbf{01}, 044 (2009),
  \eprint{0712.0743}.

\bibitem[Brigante et~al.(2008)]{Brigante:2007nu}
M.~Brigante, H.~Liu, R.~C. Myers, S.~Shenker, and S.~Yaida, \emph{Phys. Rev.}
  \textbf{D77}, 126006 (2008), \eprint{0712.0805}.

\bibitem[Buchel et~al.(2009)]{Buchel:2008vz}
A.~Buchel, R.~C. Myers, and A.~Sinha, \emph{JHEP} \textbf{03}, 084 (2009),
  \eprint{0812.2521}.

\bibitem[Baier et~al.(2008)]{Baier:2007ix}
R.~Baier, P.~Romatschke, D.~T. Son, A.~O. Starinets, and M.~A. Stephanov,
  \emph{JHEP} \textbf{0804}, 100 (2008), \eprint{0712.2451}.

\bibitem[Bhattacharyya et~al.(2008)]{Bhattacharyya:2008jc}
S.~Bhattacharyya, V.~E. Hubeny, S.~Minwalla, and M.~Rangamani, \emph{JHEP}
  \textbf{0802}, 045 (2008), \eprint{0712.2456}.

\bibitem[Kharzeev and Son(2011)]{Kharzeev:2010gr}
D.~E. Kharzeev, and D.~T. Son, \emph{Phys.Rev.Lett.} \textbf{106}, 062301
  (2011), \eprint{1010.0038}.

\bibitem[Erdmenger et~al.(2009)]{Erdmenger:2008rm}
J.~Erdmenger, M.~Haack, M.~Kaminski, and A.~Yarom, \emph{JHEP} \textbf{0901},
  055 (2009), \eprint{0809.2488}.

\bibitem[Banerjee et~al.(2011)]{Banerjee:2008th}
N.~Banerjee, J.~Bhattacharya, S.~Bhattacharyya, S.~Dutta, R.~Loganayagam,
  et~al., \emph{JHEP} \textbf{1101}, 094 (2011), \eprint{0809.2596}.

\bibitem[Son and Surowka(2009)]{Son:2009tf}
D.~T. Son, and P.~Surowka, \emph{Phys.Rev.Lett.} \textbf{103}, 191601 (2009),
  \eprint{0906.5044}.

\bibitem[Herzog et~al.(2006)]{Herzog:2006gh}
C.~Herzog, A.~Karch, P.~Kovtun, C.~Kozcaz, and L.~Yaffe, \emph{JHEP}
  \textbf{0607}, 013 (2006), \eprint{hep-th/0605158}.

\bibitem[Casalderrey-Solana and Teaney(2006)]{CasalderreySolana:2006rq}
J.~Casalderrey-Solana, and D.~Teaney, \emph{Phys.Rev.} \textbf{D74}, 085012
  (2006), \eprint{hep-ph/0605199}.

\bibitem[Gubser(2006)]{Gubser:2006bz}
S.~S. Gubser, \emph{Phys.Rev.} \textbf{D74}, 126005 (2006),
  \eprint{hep-th/0605182}.

\bibitem[Ficnar et~al.(2011)]{Ficnar:2011yj}
A.~Ficnar, J.~Noronha, and M.~Gyulassy  (2011), * Temporary entry *,
  \eprint{1106.6303}.

\bibitem[Chesler et~al.(2009)]{Chesler:2008uy}
P.~M. Chesler, K.~Jensen, A.~Karch, and L.~G. Yaffe, \emph{Phys.Rev.}
  \textbf{D79}, 125015 (2009), \eprint{0810.1985}.

\bibitem[Gubser et~al.(2008)]{Gubser:2008as}
S.~S. Gubser, D.~R. Gulotta, S.~S. Pufu, and F.~D. Rocha, \emph{JHEP}
  \textbf{0810}, 052 (2008), \eprint{0803.1470}.

\bibitem[Baier et~al.(1996)]{Baier:1996vi}
R.~Baier, Y.~L. Dokshitzer, A.~H. Mueller, S.~Peigne, and D.~Schiff,
  \emph{Nucl. Phys.} \textbf{B478}, 577--597 (1996), \eprint{hep-ph/9604327}.

\bibitem[Baier et~al.(1997)]{Baier:1996kr}
R.~Baier, Y.~L. Dokshitzer, A.~H. Mueller, S.~Peigne, and D.~Schiff,
  \emph{Nucl. Phys.} \textbf{B483}, 291--320 (1997), \eprint{hep-ph/9607355}.

\bibitem[Liu et~al.(2006)]{Liu:2006ug}
H.~Liu, K.~Rajagopal, and U.~A. Wiedemann, \emph{Phys.Rev.Lett.} \textbf{97},
  182301 (2006), \eprint{hep-ph/0605178}.

\bibitem[Arnold and Vaman(2010)]{Arnold:2010ir}
P.~Arnold, and D.~Vaman, \emph{JHEP} \textbf{1010}, 099 (2010),
  \eprint{1008.4023}.

\bibitem[Chesler and Yaffe(2011)]{Chesler:2010bi}
P.~M. Chesler, and L.~G. Yaffe, \emph{Phys.Rev.Lett.} \textbf{106}, 021601
  (2011), \eprint{1011.3562}.

\end{thebibliography}

\IfFileExists{\jobname.bbl}{}
 {\typeout{}
  \typeout{******************************************}
  \typeout{** Please run "bibtex \jobname" to optain}
  \typeout{** the bibliography and then re-run LaTeX}
  \typeout{** twice to fix the references!}
  \typeout{******************************************}
  \typeout{}
 }

\end{document}